\documentclass[12pt]{article}

\usepackage{amsmath}
\usepackage{graphicx}

\title{On the overall polarisation properties of full Poincar\'{e} beams}
\author{Dorilian Lopez-Mago\\Tecnologico de Monterrey}
\date{June 2019\\ dlopezmago(at)tec.mx}
\begin{document}
\maketitle


\section{abstract}
We analyse the polarisation properties of full Poincar\'{e} beams. We consider different configurations, such as Laguerre-Poincar\'{e}, Bessel-Poincar\'{e}, and Lambert-Poincar\'{e} beams. The former is the original Poincar\'{e} beam produced by a collinear superposition of two Laguerre-Gauss beams with orthogonal polarisations. For this configuration, we describe the Stokes statistics and overall invariant parameters. Similarly, Bessel-Poincar\'{e} beams are produced by the collinear superposition of Bessel beams with orthogonal polarisations. We describe their properties under propagation and show that they behave as a free-space polarisation attractor transforming elliptical polarisations to linear polarisations. We also propose a novel type of full Poincar\'{e} pattern, one which is generated by a Lambert projection of the Poincar\'{e} sphere on the transverse plane, and hence we call them Lambert-Poincar\'{e}. This configuration, contrary to the Laguerre-Poincar\'{e}, provides a finite region containing all polarisation states uniformly distributed on the Poincar\'{e} sphere.

\section{Introduction}

Optical beams containing all states of polarisation are known as full Poincar\'{e} (\textbf{FP}) beams~\cite{Beckley2010,Galvez2012,Shvedov(2015)}. Their polarisation properties have attracted attention in beam shaping and singular optics, where they can be used to generate complex topologies such as M\"{o}bius strips~\cite{Bauer2015}. Other studies analyse the optical forces arising from the curl of the spin angular momentum~\cite{Wang(2012)}. They can be applied in laser cutting, where the flat-top intensity is used to realise a clean cut~\cite{Han(2011)}. Potential applications have been shown for quantum communications~\cite{Fickler(2014)}, single-shot polarimetry~\cite{Sivankutty(2016)}, and polarisation speckles~\cite{Reddy(2017)}. 

Full-Poincar\'{e} beams were originally realised by Beckley et al~\cite{Beckley2010}. They are formed by a coherent and collinear superposition of two Laguerre-Gauss (LG) beams with either linear or circular orthogonal polarisations. For the circular basis, two distinct polarisation singularities can be generated, that is, lemon and star singularities. Higher-order \textbf{FP} beams were studied by Galvez et al~\cite{Galvez2012}, where superpositions of higher-order LG beams were used to produce the mapping of multiple Poincar\'{e} spheres on the transverse plane. Other type of \textbf{FP} beam include the Bessel-Poincar\'{e} studied by Shvedov et al~\cite{Shvedov(2015)}, which is formed by a superposition of Bessel beams. Similarly, there are Mathieu-Poincar\'{e} beams, which are formed by using nondiffracting Mathieu beams~\cite{Garcia(2016)}.

In this work, we describe interesting polarisation properties regarding \textbf{FP} beams. We consider the original \textbf{FP} beams formed with an LG mode basis. We call them Laguerre-Poincar\'{e} (\textbf{LP}) to distinguish them from the other structures of interest, which are Bessel-Poincar\'{e} (\textbf{BP}) and Lambert-Poincar\'{e} (\textbf{LaP}) beams. The latter is realised through a Lambert projection of the Poincar\'{e} sphere on the transverse plane. Section~\ref{Sec:StokesStatistics} describes the Stokes statistics that we use to describe the overall polarisation properties of the \textbf{FP} beams. Inspired by the work of Mart\'{i}nez-Herrero et al~\cite{Herrero2006}, we found an invariant parameter that relates the local degree of polarisation with the averaged Stokes parameters. We use this formalism to study the properties of \textbf{LP} beams in section~\ref{Sec:Laguerre}. We describe the Stokes variances according to the azimuthal and radial indices and show that higher-order \textbf{LP} beams are predominantly circularly polarised. Furthermore, we study their propagation and describe the evolution of the transverse polarisation distribution as a rotation of the Poincar\'{e} sphere. Similarly, section~\ref{Sec:Bessel} studies the propagation of \textbf{BP} beams and show that they behave as a polarisation attractor~\cite{Pitois2008}, which transforms elliptical polarisation states to linear polarisations. Finally, section~\ref{Sec:Lambert} explains the realisation of the \textbf{LaP} polarisation pattern, which provides a uniform distribution of polarisation states on the Poincar\'{e} sphere~\cite{Yan2005}.

\section{Global parameters and invariants for the characterisation of space-variant polarised beams}\label{Sec:StokesStatistics}

In this section we briefly describe the Stokes statistics in order to stablish basic formulas and notation. The Stokes parameters, $S_{j}$ with $j \equiv 0,1,2,3$, are determined by six power measurements. Each power measurement is realised after the beam passes through an ideal polariser. The Stokes parameters are defined as~\cite{Chipman2019}
\begin{equation}
\mathbf{S} = (P,Q,U,V)=(P_{H}+P_{V},P_{H}-P_{V},P_{45}-P_{135},P_{R}-P_{L}),
\end{equation}
where $P$ is equal to the total power of the beam. $Q$ is the power transmitted after a horizontal polariser ($P_{H}$) minus the power transmitted through a vertical polariser ($P_{V}$). Similarly, $U$ is the power transmitted after a $45^{\circ}$ polarizer ($P_{45}$) minus the power after a $135^{\circ}$ polariser ($P_{135}$).  Finally, $V$ measures the difference in power between the right-handed circular component ($P_{R}$) and the left-handed circular component ($P_{L}$). We label these global Stokes parameters as $P$, $Q$, $U$, $V$ to distinguish them from the space-dependent Stokes parameters defined later. 

For space-variant polarised beams, e.g. FP beams, each power measurement is equal to the spatial integration of their respective intensities. For instance
\begin{equation}
P =P_{H} + P_{V} = \int \int (I_{H}(x,y) + I_{V}(x,y))\mathrm{d}x\mathrm{d}y = \int \int I(x,y)\mathrm{d}x\mathrm{d}y,
\end{equation}
where $I(x,y)$ is the total intensity of the beam, $I_{H}(x,y)$ and $I_{V}(x,y)$ are the horizontal and vertical intensity components, respectively. Correspondingly, $I_{45}$, $I_{135}$, $I_{R}$ and $I_{L}$ are the intensity components for $P_{45}$, $P_{135}$, $P_{R}$ and $P_{L}$. The intensities are measured with a CCD camera whereas the power or flux of energy is measured using a photodetector. To keep the notation simpler, we will disregard the space dependence $(x,y)$ unless is necessary for clarity. 

We define the normalised and space-variant Stokes parameters
\begin{eqnarray}
s_{1}(x,y) &=&  (I_{H} - I_{V})/I, \label{Eq:s1} \\
s_{2}(x,y) &=&  (I_{45} - I_{135})/I, \label{Eq:s2} \\
s_{3}(x,y) &=&  (I_{R} - I_{L})/I. \label{Eq:s3}
\end{eqnarray}
Inspired by the work of Mart\'{i}nez-Herrero et al~\cite{Herrero2006}, we consider in our analyses the weighted average of the normalised Stokes parameters, which are given by
\begin{equation}
\langle s_{j} \rangle = \frac{\int \int s_{j}(x,y) I (x,y) \mathrm{d}x \mathrm{d}y}{P},
\end{equation}
with $j\equiv 1,2,3$. Notice that $I(x,y)$ in this definition is equivalent to a density function. In addition, the expected value of the variance  $\sigma_{j} = (s_{j} - \langle s_{j} \rangle)^{2} $ is defined as
\begin{equation} \label{Eq:sigma}
\langle \sigma_{j} \rangle = \frac{\int \int (s_{j}(x,y) - \langle s_{j} \rangle)^{2} I(x,y)  \mathrm{d}x \mathrm{d}y}{P}.
\end{equation}
We notice that Mart\'{i}nez-Herrero et al used these definitions to characterise space-variant polarised beams in terms of one or two Stokes parameters~\cite{Herrero2006,Herrero(2008)}. However, in this work, we show that considering the three Stokes parameters provide information regarding the distribution of polarisation states over the Poincar\'{e} sphere. 

In what follows we will repeatedly use the definition
\begin{equation}
\langle A \rangle = \frac{\int \int A(x,y) I(x,y)\mathrm{d}x\mathrm{d}y}{P},
\end{equation}
with $A \equiv s_{j},s_{j}^{2},\sigma_{j},\mathrm{etc}$. By expanding equation~\ref{Eq:sigma}, we can write $\langle \sigma_{j} \rangle$ in the alternative form
\begin{equation}
\langle \sigma_{j} \rangle = \langle s_{j}^{2} \rangle - \langle s_{j} \rangle^{2}.
\end{equation}
We now consider the contribution from the three Stokes parameters. Let us define 
\begin{equation} \label{Eq:Omega}
\Omega = \sum_{j=1}^{3} \langle \sigma_{j} \rangle = \sum_{j=1}^{3}\left( \langle s_{j}^{2} \rangle - \langle s_{j} \rangle^{2} \right),
\end{equation}
as the total variance of polarisation states. We observe that the first term on the right-hand side,
\begin{equation}
\sum_{j=1}^{3} \langle s_{j}^{2} \rangle = \frac{\int \int (s_{1}^{2}+s_{2}^{2}+s_{3}^{2})I(x,y) \mathrm{d}x\mathrm{d}y}{P},
\end{equation}
is equal to the expected value of the squared of the local degree of polarisation (LDoP), defined as (cf.~\cite{Herrero2006})
\begin{equation}
\mathrm{LDoP} (x,y) = \sqrt{s_{1}^{2}(x,y) + s_{2}^{2}(x,y)+s_{3}^{2}(x,y)}.
\end{equation}
Thus we have that
\begin{equation}
\sum_{j=1}^{3} \langle s_{j}^{2} \rangle = \langle \mathrm{LDoP}^{2} \rangle.
\end{equation}
 Then, the addition of the squared of the average values, turns out to be equal to the squared of the global degree of polarisation (DoP), i.e.
\begin{equation}
\sum_{j=1}^{3} \langle s_{j} \rangle^{2} = \frac{Q^{2} + U^{2} + V^{2}}{P^{2}} = \mathrm{DoP}^2.
\end{equation}
Therefore, equation~\ref{Eq:Omega} can be written as
\begin{equation}
\Omega = \langle \mathrm{LDoP}^{2} \rangle - \mathrm{DoP}^2. 
\end{equation}
By knowing that the global DoP is invariant under propagation through non-polarising optical elements~\cite{Chipman2019}, it is convenient to write the previous equation as
\begin{equation} \label{Eq:invariant}
 \mathrm{DoP}^2 = \langle \mathrm{LDoP}^{2} \rangle - \Omega. 
\end{equation}
Equation~\ref{Eq:invariant} is the first important result of this work. It helps to characterise space-variant polarised beams. The DoP describes the overall state of polarisation whereas the variances describe the spread of the polarisation states on the principal Stokes axes. The DoP has an intuitive interpretation. If we graph the polarisation states on the Poincar\'{e} sphere, the DoP resembles the concept of center of mass. Therefore, if the polarisation states span the Poincar\'{e} sphere or one of its great circles, it results that $\mathrm{DoP}=0$, as is the case for \textbf{FP} and cylindrical vector beams. The variances give information about the predominant state of polarisation according to the intensity of the beam. For example, values of $\langle \sigma_{3} \rangle > \langle \sigma_{1,2} \rangle$ means that the beam is circularly polarised in regions where the intensity is more significant. 

\begin{figure}[htbt]
\centering
  \includegraphics[width=8 cm]{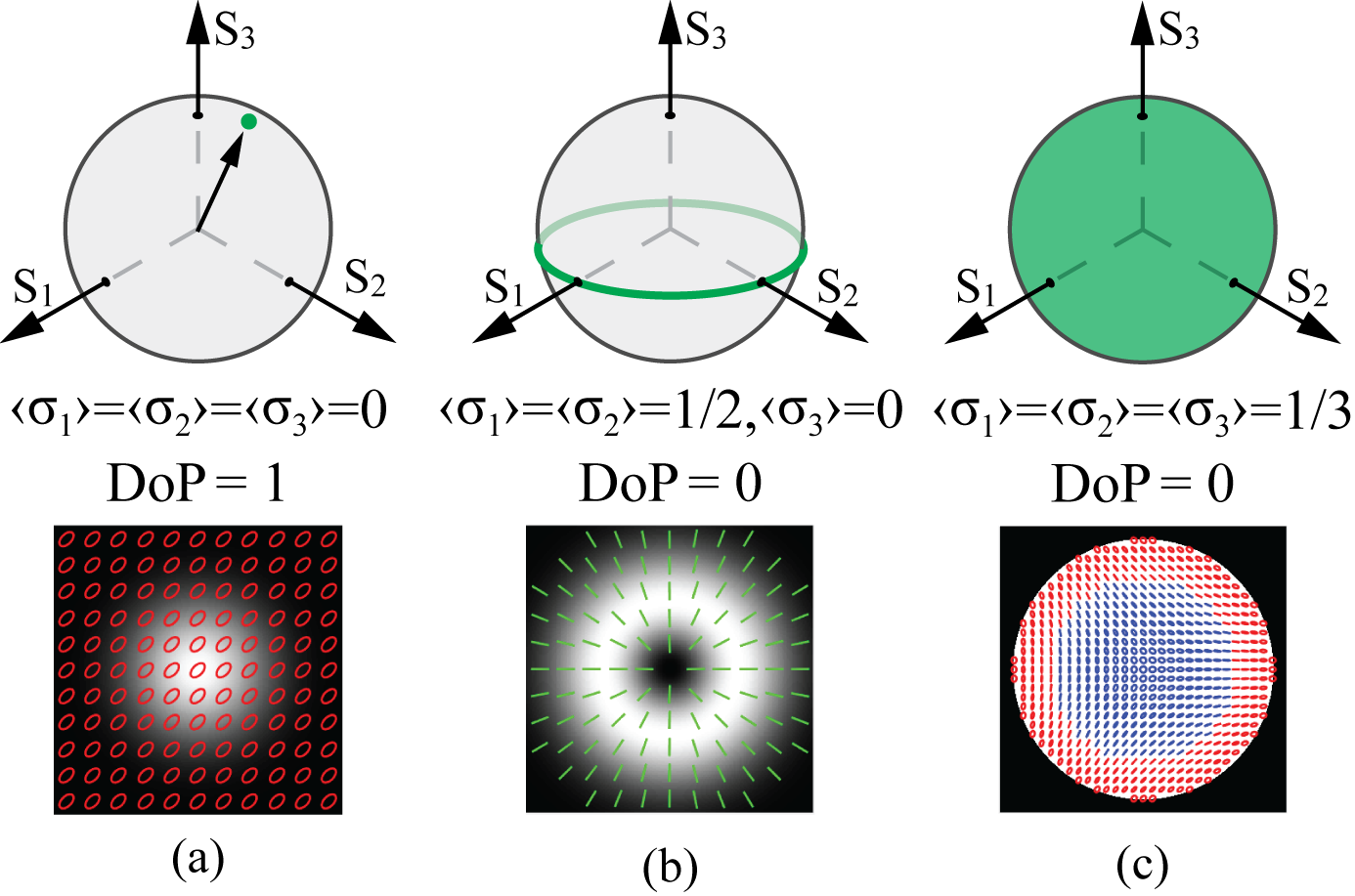}
  \caption{(Colour online). Stokes statistics for (a) a unirformly polarised beam, (b) a radially polarised beam, and (c) a Lambert-Poincar\'{e} beam. DoP stands for degree of polarisation. $\langle \sigma_{1,2,3} \rangle$ are the averaged Stokes variances. They satisfy equation~\ref{Eq:sigmasequal1}. The gray colormap is proportional to the intensity. Red (blue) ellipses are polarisation states with right (left) handedness, whereas the green lines are linear polarisation states. The sphere shows the polarisation states of the transverse plane on the Poincar\'{e} sphere.}
  \label{fig:1}
\end{figure}

In the examples that follow, we consider temporally and spatially coherent light beams (e.g. monochromatic laser beams). It means that the light is locally polarised and hence $\langle \mathrm{LDoP} \rangle = 1$. Therefore $\mathrm{DoP}^2+\Omega = 1$, which is written as
\begin{equation}
\langle \sigma_{1} \rangle+\langle \sigma_{2} \rangle+\langle \sigma_{3} \rangle = 1 - \mathrm{DoP}^2. \label{Eq:sigmasequal1}
\end{equation}
Figure~\ref{fig:1} shows examples of the application of the previous equation. We can see that for homogeneously polarised beams  $\mathrm{DoP} = 1$ and hence $\langle \sigma_{1,2,3}\rangle = 0$. For \textbf{FP} and cylindrical vector beams $\mathrm{DoP}=0$. Cylindrical vector beams span the equator of the Poincar\'{e} sphere, meaning that $\langle \sigma_{3} \rangle = 0$ and $\langle \sigma_{1}\rangle =\langle \sigma_{2} \rangle = 1/2$ (see figure~\ref{fig:1}(b)). Figure~\ref{fig:1}(c) shows a \textbf{LaP} beam which uniformly covers the Poincar\'{e} sphere and hence  $\langle \sigma_{1}\rangle = \langle \sigma_{2} \rangle =\langle \sigma_{3}\rangle = 1/3$.

\section{Polarisation properties of Laguerre-Poincar\'{e} beams}\label{Sec:Laguerre}

Under the paraxial approximation an LG beam is written in a cylindrical coordinate system $\mathbf{r}=(r,\phi,z)$ as~\cite{Plick2015}
\begin{eqnarray}
\mathrm{LG}_{n}^{m} (r,\phi,z) &=& \sqrt{\frac{2(n!)}{\pi (n+|m|)!}}\frac{1}{w_{z}}\left(\frac{\sqrt{2}r}{w_{z}} \right)^{|m|} \exp \left(- \frac{r^{2}}{w_{z}^{2}}\right)\times \nonumber \\  & & \mathrm{L}_{n}^{|m|}\left( \frac{2r^{2}}{w_{z}^{2}} \right) \exp\left( i \frac{kr^{2}}{2R_{z}} \right) \exp \left( - i \Phi \right) \exp(i m \phi), \label{Eq:LG}
\end{eqnarray}
where $n$ and $m$ are the radial and azimuthal quantum numbers, respectively, and they also determine the order $n$ and degree $m$ of the generalised Laguerre polynomial $\mathrm{L}_{n}^{|m|}$. Moreover,
\begin{eqnarray}
z_{R} &=& \frac{1}{2} k w_{0}^{2},\\
w_{z}&=& w_{0} \sqrt{1+(z/z_{R})^{2}},\\
R_{z}&=&\frac{z^{2} +z_{R}^{2} }{z},\\
\Phi &=& \left(2n+|\ell|+1\right)\mathrm{arctan}(z/z_{R}), \label{Eq:Gouy}
\end{eqnarray}
where $z_{R}$ is the Rayleigh range, $k=2\pi/\lambda$ is the wave number, $w_{z}$ is the beam waist (so $w_{0}$ is the beam waist at $z=0$), $R_{z}$ is the radius of curvature, and $\Phi$ is the Gouy phase. Furthermore, equation~\ref{Eq:LG} is normalised such that $\int \int |\mathrm{LG}|^{2} \mathrm{d}x\mathrm{d}y = 1$.

Laguerre-Poincar\'{e} (\textbf{LP}) beams are produced with a coherent and collinear superposition of LG beams with orthogonal polarisations. We  consider a circular polarisation basis and use the unit vectors $\mathbf{\hat{c}}_{R} = (\mathbf{\hat{x}} - i \mathbf{\hat{y}})/\sqrt{2}$ and $\mathbf{\hat{c}}_{L} = (\mathbf{\hat{x}} + i \mathbf{\hat{y}})/\sqrt{2}$, which correspond to the right-handed and left-handed circular polarisations, respectively. Thus, an \textbf{LP} beam with circular polarisation basis is given by~\cite{Galvez2012}
\begin{equation} \label{Eq:LPbeams}
\mathbf{LP} = \mathrm{LG}_{n_R}^{m_R}\mathbf{\hat{c}}_{R} + \mathrm{LG}_{n_L}^{m_L}\mathbf{\hat{c}}_{L},
\end{equation}
where we consider that both components can have different quantum numbers. Another polarisation basis can be used to generate \textbf{LP} beams, whose difference is observed in the transverse distribution of the polarisation states. The factor of $1/\sqrt{2}$ ensures that  $\int \int |\mathbf{LP}|^{2} \mathrm{d}x\mathrm{d}y = 1$. Notice that equation~\ref{Eq:LPbeams} does not span all polarisation states for all combinations of $m_{R},m_{L}$ and $n_{R},n_{L}$. The cases where one of the components has $m= n=0$ and the other component has $m=1,n=0$ correspond to the first-order \textit{LP} beams~\cite{Beckley2010}.  Higher-order \textbf{LP} beams are generated, for example, with $m_{R}=n_{R}=0$ and $m_{L} >1,n_{L}=0$~\cite{Han(2011)}. Furthermore, cylindrical vector beams correspond to $m_{R}=-m_{L}$ and $n_{R}=n_{L}=0$~\cite{Zhan(2009)}.

\begin{figure}[htbt]
\centering
  \includegraphics[width=8 cm]{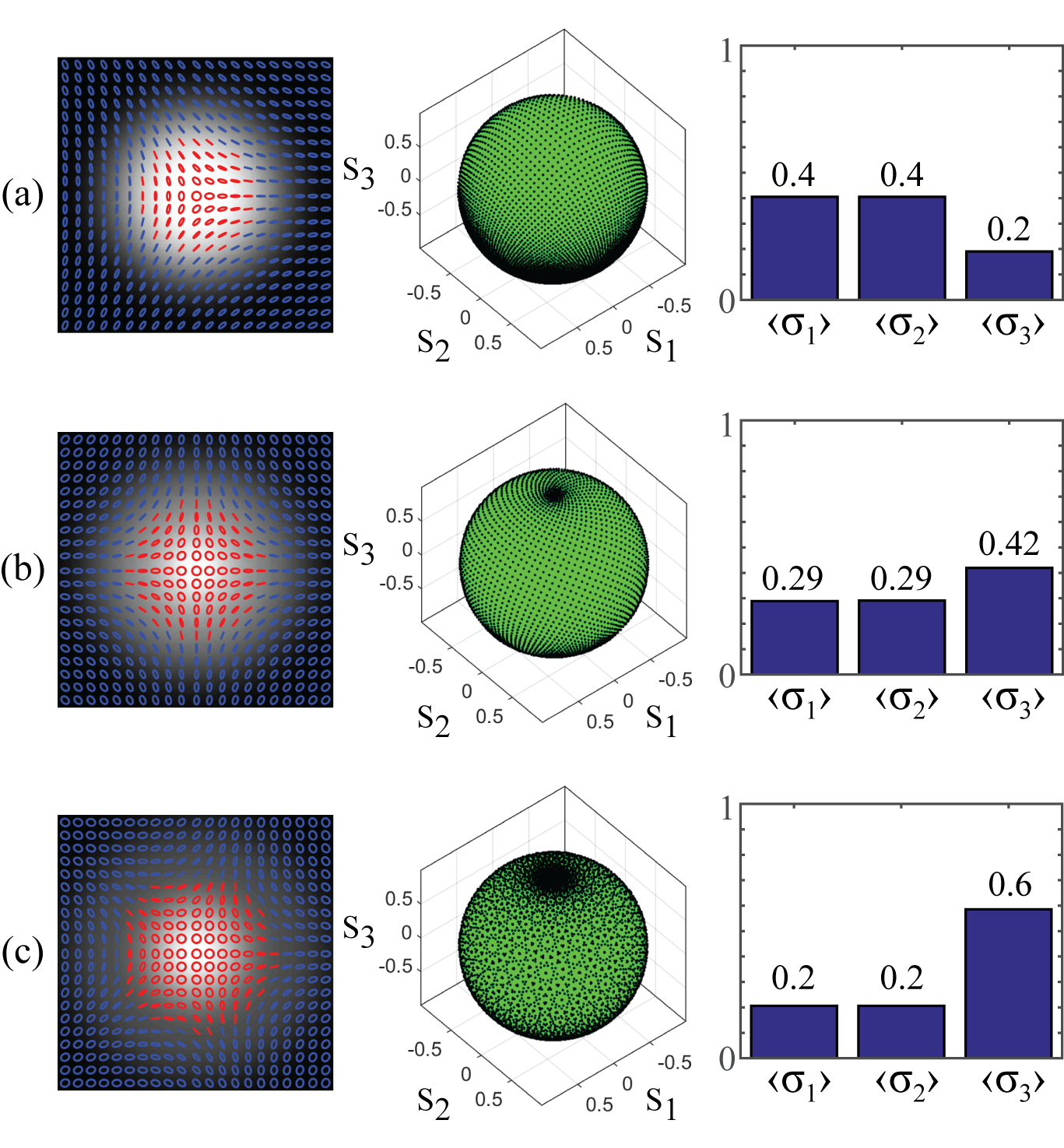}
  \caption{(Colour online). Stokes variances for \textbf{LP} beams (see equation~\ref{Eq:LPbeams}) with $n_{R}=n_{L}=0$ and (a) $(m_{R},m_{L})=(0,1)$, (b) $(m_{R},m_{L})=(0,2)$, (c) $(m_{R},m_{L})=(0,3)$. The transverse profiles have dimensions of $3w_{0}\times 3w_{0}$.}
  \label{fig:2}
\end{figure}

Figure~\ref{fig:2} shows the properties for different \textbf{LP} beams with both polarisation components carrying a zero radial quantum number (i.e. $n_{R}=n_{L}=0$). First, second and third columns show the transverse polarisation distribution, the polarisation states visualised on the Poincar\'{e} sphere and the Stokes variances (see equation~\ref{Eq:sigma}), respectively. The calculations were done at the plane $z=0$ with $w_{0}=4$ mm and $\lambda = 633$ nm. For the transverse polarisation distributions, the gray colormap represents the intensity, whereas the red (blue) ellipses represent polarisation states with right (left) handedness. The graph of the polarisation states on the Poincar\'{e} sphere is realised by mapping each pixel of the transverse plane on a unit sphere using the normalised Stokes parameters (see equations~\ref{Eq:s1} to~\ref{Eq:s3}). 

Figure~\ref{fig:2}(a) shows a first-order \textbf{LP}  beam with a star singularity ($m_{R}=0,m_{L}=1$). From the transverse pattern, notice that the north hemisphere, the one containing right-handed polarisation states, is located where the intensity of the beam is significant. Specifically, the north hemisphere has about 90 percent of the total power (it was calculated at the radial position $r=w_{0}/\sqrt{2}$ that satisfies $s_{3} = 0$, which is the location of the equator). Therefore, the left-handed hemisphere covers a region with practically null intensity. This causes that $\langle \sigma_{3} \rangle$ is less significant than $\langle \sigma_{1} \rangle$ and $\langle \sigma_{2} \rangle$, as shown in the third column. 

Figures~\ref{fig:2}(b) and~\ref{fig:2}(c) show the results for higher-order \textbf{LP} beams with $(m_{R},m_{L})=(0,1)$ and $(m_{R},m_{L})=(0,2)$ (cf.~\cite{Otte2016}). Since the LG beam radius increases with the azimuthal quantum number due to the factor $(\sqrt{2}r/w_{z})^{|m|}$ in equation~\ref{Eq:LG}, the overlap between the two LG components decreases for high values of $m_{R}$. Therefore, the circular polarisation states are dominant for higher-order \textbf{LP} beams. Notice the accumulation of polarisation states on the poles of the Poincar\'{e} sphere. Similarly, the value of $\langle \sigma_{3} \rangle$ becomes more significant than $\langle \sigma_{1} \rangle$ and $\langle \sigma_{2} \rangle$. 

Due to the circular polarisation basis, the results discussed in the previous paragraphs satisfy $\langle \sigma_{1} \rangle=\langle \sigma_{2} \rangle$. Furthermore, notice that in all cases $\sum \langle \sigma_{j} \rangle=1$ is satisfied. Of course, we can infer similar conclusions according to the polarisation basis that we use. For example, if we use a linear polarisation basis $\mathbf{\hat{x}},\mathbf{\hat{y}}$, it is expected that $\langle \sigma_{2} \rangle=\langle \sigma_{3} \rangle$, and the polarisation states on the Poincar\'{e} sphere will accumulate on $s_{1} = \pm 1$ for higher-order modes (and therefore $\langle \sigma_{1} \rangle>\langle \sigma_{2,3} \rangle$ ).  

\begin{figure}[htbt]
\centering
  \includegraphics[width=8 cm]{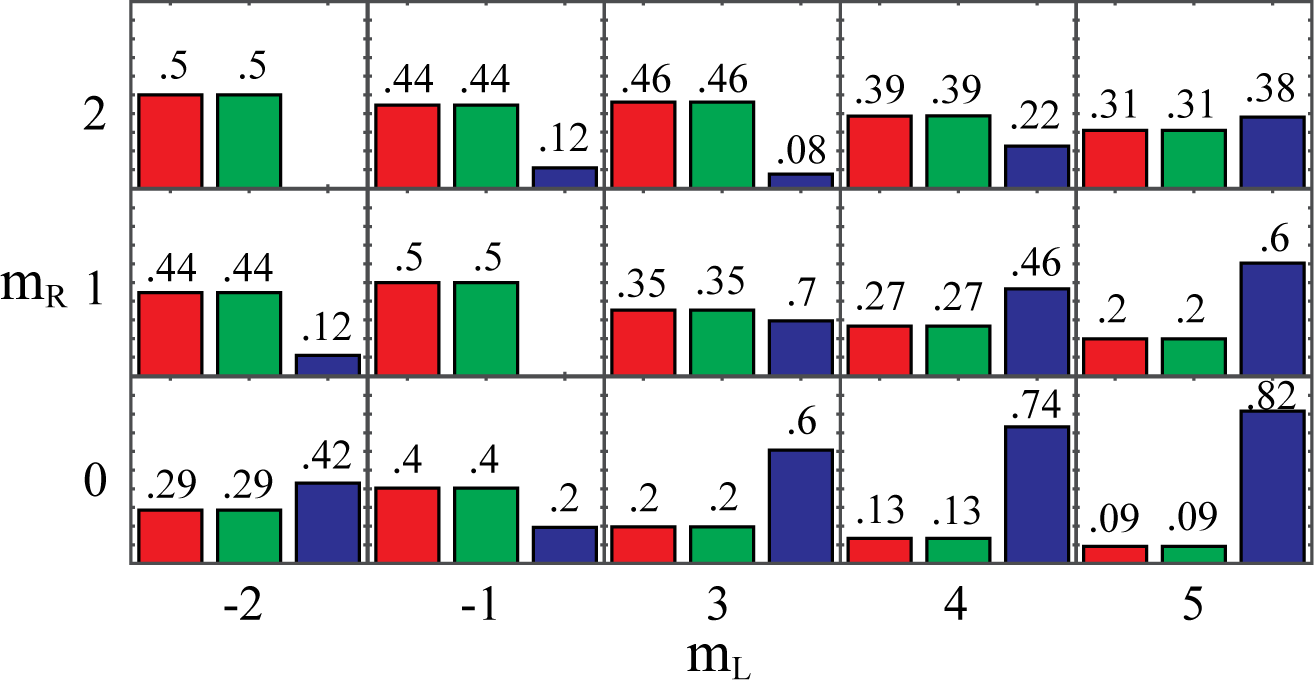}
  \caption{(Colour online). Stokes variances for different combinations of \textbf{LP} beams with $n_{R}=n_{L}=0$, according to~\ref{Eq:LPbeams}. Red, Green, Blue bars correspond to $\langle \sigma_{1} \rangle, \langle \sigma_{2} \rangle, \langle \sigma_{3} \rangle$, respectively.}
  \label{fig:3}
\end{figure}

To extend our previous analysis, figure~\ref{fig:3} shows more combinations of $m_{R},m_{L}$ including negative values. The cases with $m_{R}=-m_{L}$, where $\langle \sigma_{1} \rangle=\langle \sigma_{2} \rangle=1/2$ correspond to cylindrical vectors beams. The results confirm that $\langle \sigma_{1} \rangle=\langle \sigma_{2} \rangle$ and the predominant variance is $\langle \sigma_{3} \rangle$ for values of $m_{R}>>m_{L}$ (or viceversa $m_{L}>>m_{R}$).

\begin{figure}[htbt]
\centering
  \includegraphics[width=12 cm]{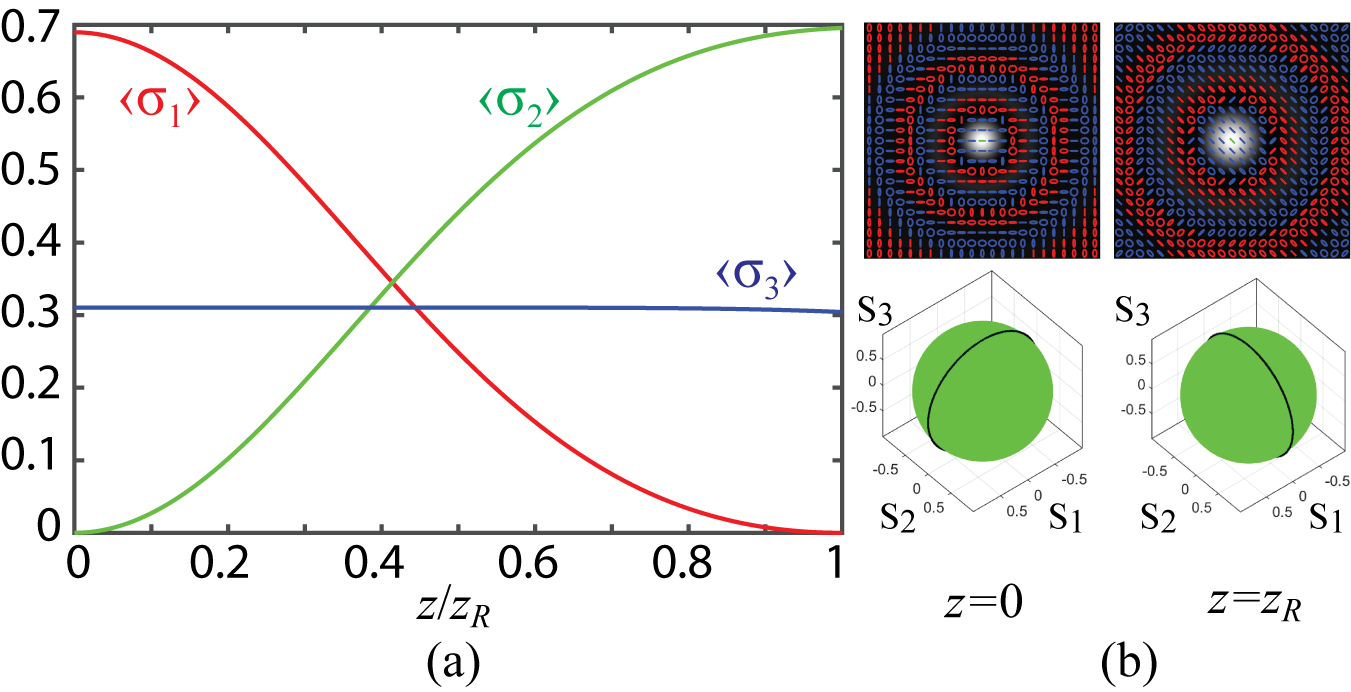}
  \caption{(Colour online). (a) Stokes variances during the propagation of a \textbf{LP} beam with $(n_{R},n_{L})=(3,2)$ and $m_{R}=m_{L}=0$. Red, gree, blue curves correspond to $\langle \sigma_{1} \rangle,\langle \sigma_{2} \rangle,\langle \sigma_{2} \rangle$, respectively. (b) Transverse polarisation distribution and its representation on the Poincar\'{e} sphere for the initial ($z=0$) and final plane ($z=z_{R}$).}
  \label{fig:4}
\end{figure}

Next we study the propagation effects on the Stokes statistics. It has been shown that under propagation the first-order \textbf{LP} beams experience a rotation of their polarisation states equivalent to a rotation of the Poincar\'{e} sphere. This rotation depends on the polarisation basis and is due to the Gouy phase (cf. equation~\ref{Eq:Gouy}). For circular polarisation basis the rotation is with respect to the $s_{3}$ axis (which are the poles of the Poincar\'{e}  sphere). Since  $\langle \sigma_{1} \rangle=\langle \sigma_{2} \rangle$ we can expect that this rotation does not change the values of the variances, which is only true for \textbf{LP} beams that cover the Poincar\'{e} sphere or its equator. Thus, the rotation does not change the polarisation state distribution on the sphere. Nevertheless, there are cases where the polarisation states cover a great circle of the Poincar\'{e} sphere other than the equator. For these cases, we can infer that the rotation will change the  values of $\langle \sigma_{1} \rangle$ and $\langle \sigma_{2} \rangle$. To show this effect, figure~\ref{fig:4} shows the Stokes variances for an \textbf{LP} beam with azimuthal quantum numbers $m_R=m_L=0$ and radial numbers $n_{R}=3$ $n_{L}=2$. The plot shows the variances as a function of $z$ up to one Rayleigh distance. We confirm that $\langle \sigma_{3} \rangle$ remains constant on propagation, but $\langle \sigma_{1} \rangle$ and $\langle \sigma_{2} \rangle$ oscillate with a spatial period equal to $2z_{R}$. As shown in figure~\ref{fig:4}(b), the polarisation states at $z=0$ cover the great circle that passes through $s_{1}=\pm 1$ and $s_{3}=\pm 1$. During propagation the states rotates with respect to the $s_{3}$ axis and at $z=z_{R}$ the states cover the great circle that passes through $s_{2}=\pm 1$ and $s_{3}=\pm 1$.

We can summarise the observations for \textbf{LP} beams as follows: (i) cases with $n_{R}=n_{L}=0$ and $|m_{R}| \neq |m_{L}|$ correspond to higher-order \textbf{LP} beams that span all polarisation states. For circular polarisation basis  $\langle \sigma_{1} \rangle=\langle \sigma_{2} \rangle$ and $\langle \sigma_{3} \rangle$ is predominant if $||m_{R}|-|m_{L}||>>1$. (ii) Higher-order cylindrical vector beams are given with $n_{R}=n_{L}=0$ and $m_{R} = -m_{L}$. They satisfy $\langle \sigma_{1} \rangle=\langle \sigma_{2} \rangle=1/2$ and $\langle \sigma_{3} \rangle = 0$. (iii) For $m_{R}=m_{L}$ and $n_{R} \neq n_{L}$ the polarisation states cover a great circle different than the equator. Under propagation, the variances for the cases (i) and (ii) remain constant, but for (iii) the variances $\langle \sigma_{1} \rangle,\langle \sigma_{2} \rangle$ oscillate while $\langle \sigma_{3} \rangle$ remains constant. We remark that in all cases $\mathrm{DoP}=0$.

\section{Polarisation properties of Bessel-Poincar\'{e} beams}\label{Sec:Bessel}

Here we describe the polarisation properties of \textbf{BP} beams. The spatial modes that we use are paraxial Bessel-Gauss beams instead of nondiffracting Bessel beams. This is with the purpose of describing a beam profile closely related to an experimental realisation. Bessel-Gauss (BG) are described by the expression~\cite{Gutierrez2005}
\begin{equation}
\mathrm{BG}_{m} = \exp\left(-i \frac{k_{t}^{2}}{2k}\frac{z}{\mu}\right) \mathrm{GB}(\mathbf{r}) J_{m}\left(\frac{k_{t}r}{\mu} \right)\exp(i m \phi), 	
\end{equation}
where
\begin{equation}
\mathrm{GB}(\mathbf{r}) = \frac{\exp(ikz)}{\mu}\exp\left( - \frac{r^{2}}{\mu w_{0}^{2}} \right),
\end{equation}
is a Gaussian beam that apodises the Bessel function $J_{m}$. The index $m$ is the azimuthal quantum number that describes the amount of orbital angular momentum. Furthermore,
\begin{eqnarray}
\mu &=& 1 + iz/z_{R}, \\
\gamma &=& k_{t} w_{0}/2.
\end{eqnarray}
where $z_{R} = k w_{0}^{2}/2$ is tje Rayleigh distance for the GB apodization. However, the distance where the beam behaves as a nondiffracting beam is given by  
\begin{equation}
z_{max} = z_{R} / \gamma. 
\end{equation}
The transverse wavevector $k_{t}$ is related to $z_{max}$ as $k_{t} = \frac{w_{0}k}{z_{max}}$.

A first-order \textbf{BP} beam in a circular polarisation basis reads as
\begin{equation}
\mathbf{BP} = \frac{1}{\sqrt{2}}\left( \mathrm{BG}_{0}\mathbf{\hat{c}}_{R} + \mathrm{BG}_{1}\mathbf{\hat{c}}_{L} \right). \label{Eq:BP}
\end{equation}
This configuration spans the Poincar\'{e} sphere and therefore its $DoP=0$. The variances at $z=0$ are equal to $\langle \sigma_{1} \rangle = \langle \sigma_{2} \rangle = 0.24$ and $\langle \sigma_{3} \rangle =0.52$. An interesting difference between \textbf{BP} and \textbf{LP} beams appears on propagation. We notice that for \textbf{LP} beams, the transverse polarisation distribution on propagation changes according to a rigid rotation of the Poincar\'{e} sphere (see section~\ref{Sec:Laguerre}). It is known that the far-field transverse amplitude of a BG beam corresponds to a ring of intensity. In other words, the angular spectrum of a Bessel beam is a ring in $k$ space~\cite{Durnin1987}. Since the ring radius is independent of the azimuthal quantum number~\cite{Vaity2015}, the far-field intensity distribution of both components of equation~\ref{Eq:BP} are the same. An equal amplitude superposition means that the resulting polarisation pattern contains linear polarisation states. Therefore, during the propagation of a \textbf{BP} beam, the elliptical polarisation states evolve to linear polarisations. 

\begin{figure}[htbt]
\centering
  \includegraphics[width=12 cm]{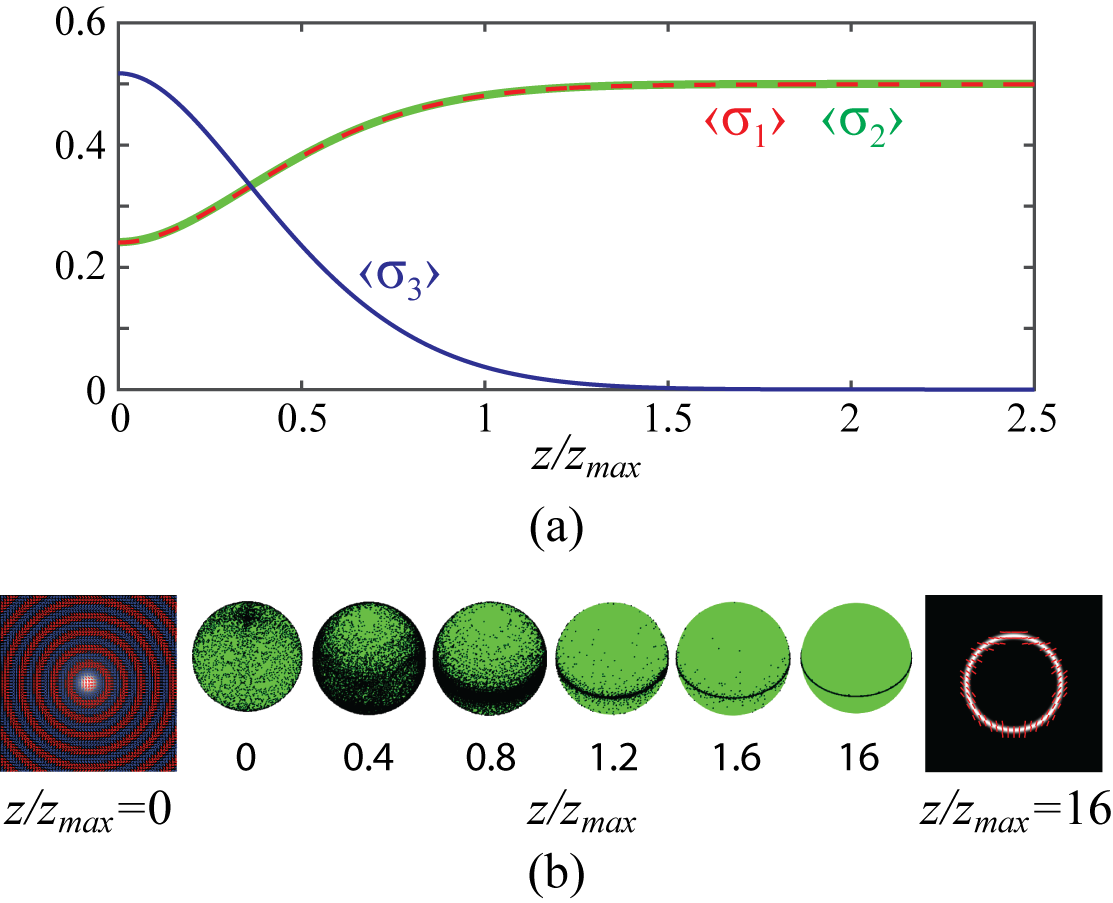}
  \caption{(Colour online). (a) Stokes variances during propagation of a \textbf{BP} beam according to equation~\ref{Eq:BP} with $(m_{R},m_{L})=(0,1)$. (b) Transverse polarisation distribution and its representation on the Poincar\'{e} sphere at different $z$ planes.}
  \label{fig:5}
\end{figure}

Figure~\ref{fig:5} shows the variances as a function of the propagation distance $z$ for a first-order \textbf{BP} beam. In the calculations we use $\lambda=633$ nm, $w_{0}=2$ mm, and $k_{t}=k \sin(0.1^{\circ})$ with $k=2\pi/\lambda$. The results confirm our previous comments. Initially, the polarisation states cover the full Poincar\'{e} sphere. During propagation, however, the polarisation states are attracted to the equator. This behaviour is compared to a polarisation attractor in optical fibres, where the states are attracted to a particular point or line on the sphere~\cite{Guasoni2014}. Notice that at all planes the $DoP=0$ and $\langle \sigma_{1} \rangle = \langle \sigma_2 \rangle$ due to the circular polarisation basis. However, contrary to the \textbf{LP} beams, the value of $\langle \sigma_3 \rangle$ is not constant during propagation. It turns out that, since the polarisation states are attracted to the equator, the value of  $\langle \sigma_3 \rangle=0$ in the far field and $\langle \sigma_{1} \rangle = \langle \sigma_2 \rangle=1/2$. Therefore, the final distribution is a particular case of a vector beam. In the example of figure~\ref{fig:5}(b) the far-field polarisation distribution is similar to a fractional vector beam of index $1/2$~\cite{Moreno2016}. 

\section{Polarisation properties of Lambert Poincar\'{e} patterns}\label{Sec:Lambert}

Here, we show a novel type of \textbf{FP} pattern that covers the Poincar\'{e} sphere with a uniform distribution of polarisation states on the sphere. The \textbf{LP} beams have been described as a stereographic projection of the Poincar\'{e} sphere on the transverse plane. Nonetheless, this projection maps one of the hemispheres to a region of infinite extension, and thus they are not physically realisable. Instead, we propose a Lambert projection to map the polarisation states on a \textit{finite} region. Of course, if the field is confined in a finite region, it will diffract during propagation. However, there are several methods to create this pattern in the image plane using spatial light modulators~\cite{Liu2012,Perez-Garcia2017,Liu2018}.

\begin{figure}[htbt]
\centering
  \includegraphics[width=12 cm]{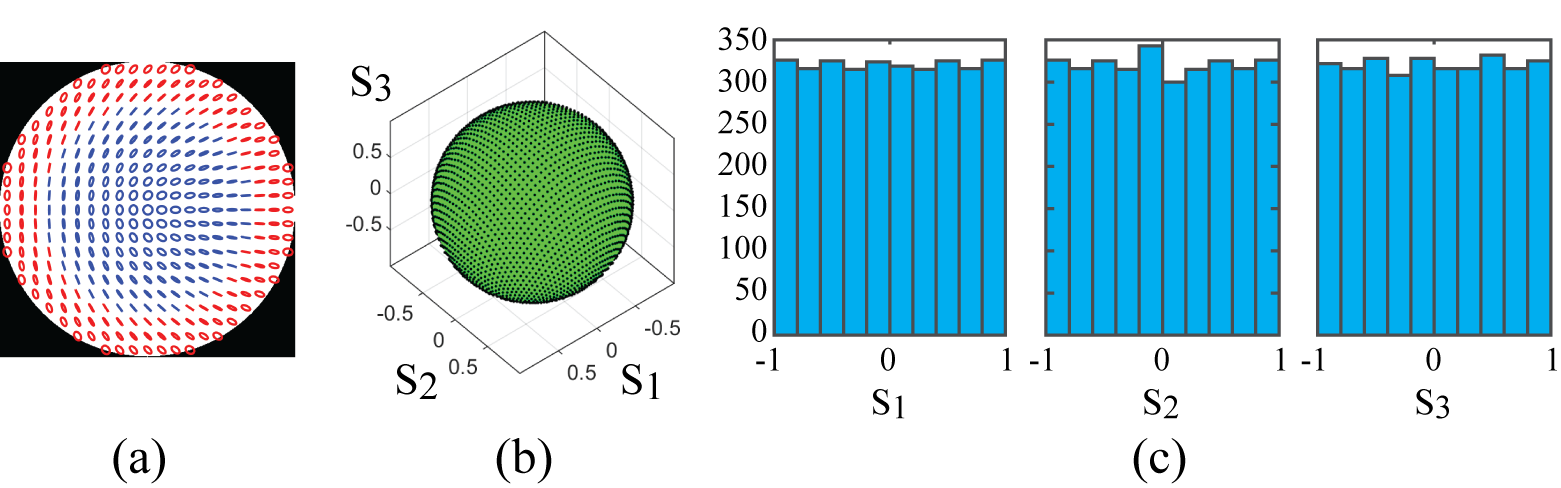}
  \caption{(Colour online). First-order Lambert-Poincar\'{e} pattern (see equation~\ref{Eq:Lambert}). (a) Polarisation distribution, (b) Transverse polarisation states represented on the Poincar\'{e} sphere, and (c) Histogram showing the distribution of the Stokes parameters.}
  \label{fig:6}
\end{figure}

We define a first-order Lambert-Poincar\'{e} pattern using circular polarisation components as
\begin{equation}
\mathbf{LaP} = (r/2) \exp(i \phi)\, \mathbf{\hat{c}}_{R} + \sqrt{1-(r/2)^{2}}\, \mathbf{\hat{c}}_{L}, \label{Eq:Lambert}
\end{equation}
where $0\leq \phi \leq 2\pi$ is the azimuthal coordinate and $r$ is a dimensionless radial coordinate confined in the range $0\leq r \leq 2$. Certainly,  higher-order \textbf{LaP} beams can be created by simply adding a larger azimuthal phase variation. 

Using the definition for the Stokes parameters in terms of circular polarisation components,
\begin{eqnarray}
s_{0} &=& |E_{R}|^{2} + |E_{L}|^{2},\\
s_{1} &=& 2 \mathrm{Re} \left(E_{R}^{\ast}E_{L} \right)/s_{0},\\
s_{2} &=& - 2 \mathrm{Im} \left(E_{R}^{\ast}E_{L} \right)/s_{0},\\
s_{3} &=&  \left(|E_{R}|^{2} - |E_{L}|^{2} \right)/s_{0},
\end{eqnarray}
we can show that $s_{0}=1$ and
\begin{eqnarray}
s_{1} &=& \sqrt{1-(r/2)^{2}}\, r \cos\phi, \\
s_{2} &=& \sqrt{1-(r/2)^{2}}\, r \sin\phi, \\
s_{3} &=& \frac{r^{2}}{2}-1.
\end{eqnarray}
Using the previous results it is straightforward to show that $P=4\pi$,  $\mathrm{DoP}=0$ and $\langle \sigma_{1} \rangle = \langle \sigma_{2} \rangle=\langle \sigma_{3} \rangle = 1/3$. 

An important property of the \textbf{LaP} pattern is that the polarisation states cover the Poincar\'{e} sphere with a uniform distribution~\cite{Yan2005,Yao2012}. Figure~\ref{fig:6}(a) shows the transverse polarisation pattern of a first-order \textbf{LaP} containing a lemon singularity (see equation~\ref{Eq:Lambert}). Figure~\ref{fig:6}(b) shows the polarisation states on the the Poincar\'{e} sphere. To show that the polarisation states are uniformly distributed on the sphere, figure~\ref{fig:6}(c) shows the histogram of polarisation states for the Stokes parameters. The histogram for each Stokes component uses $10$ equally spaced bins between $-1$ and $1$. In our calculations we use a grid of $64 \times 64$ points. The results confirm that the polarisation states are uniformly distributed on the Poincar\'{e} sphere.

\section{Discussion and conclusions}
We have studied the polarisation properties of full Poincar\'{e} beams according to the Stokes statistics developed in section~\ref{Sec:StokesStatistics}. Through the use the Stokes variances we have characterised Laguerre-, Bessel- and Lambert- Poincar\'{e} patterns. When using a circular polarisation basis, higher-order \textbf{LP} beams showed specific properties according to the radial and azimuthal quantum numbers of its components. On the one hand, equal azimuthal quantum numbers produce a polarisation pattern that covers one great circle of the Poincar\'{e} sphere and its Stokes variances change under propagation. On the other hand, dissimilar azimuthal quantum numbers produce full Poincar\'{e} patterns whose Stokes statistics are propagation invariant. We also showed that \textbf{BP} beams behave as free-space polarisation attractors, converting elliptical polarisation states to linear polarisation states. In terms of its Stokes statistics, we showed that the variances reach stable values at the far field, where the polarisation pattern is similar to a cylindrical vector beam of fractional order. Therefore, we have shown a method to generate fractional vector beams by propagating a superposition of integer-order Bessel beams. Furthermore, we have presented a \textbf{LaP} polarisation pattern whose polarisation states uniformly cover the Poincar\'{e} sphere. This is demonstrated with the histogram of the Stokes parameters which show a flat distribution. The experimental realisation and further properties of \textbf{LaP} patterns are the subject of future work. Finally, we emphasize that the aforementioned Stokes variances can be obtained with current procedures to measure the Stokes parameters. Therefore, they can be used as figures of merit to characterise space-variant polarised beams.

\section*{Acknowledgements}

We acknowledge support from Consejo Nacional de Ciencia y Tecnolog\'{i}a (CONACYT) through the grants: 257517, 280181, 293471, 295239, and APN2016-3140. I also acknowledge helpful discussions with Julio C. Guti\'{e}rrez-Vega.



\begin{thebibliography}{99}
\bibitem{Beckley2010}  Beckley A M, Brown T G and Alonso M A 2010 Full Poincar\'{e} beams  \textit{Opt. Express} \textbf{18} 10777--85
\bibitem{Galvez2012} Galvez E J, Khadka S, Schubert W H and Nomoto S 2012 Poincar\'{e}-beam patterns produced by nonseparable superpositions of Laguerre-Gauss and polarization modes of light \textit{Appl. Opt.} \textbf{51} 2925--34
\bibitem{Shvedov(2015)} Shvedov V, Karpinski P, Sheng Y, Chen X, Zhu W, Krolikowski W and Hnatovsky C 2015 Visualizing polarization singularities in Bessel-Poincar\'{e} beams \textit{Opt. Express} \textbf{23} 12444--53
\bibitem{Bauer2015} Bauer T et al 2015 Observation of optical polarization M\"{o}bius strips \textit{Science} \textbf{347} 964--6
\bibitem{Wang(2012)} Wang L G 2012 Optical forces on submicron particles induced by full Poincar\'{e} beams \textit{Opt. Express} \textbf{20} 20814--26
\bibitem{Han(2011)} Han W, Cheng W and Zhan Q 2011 Flattop focusing with full Poincar\'{e} beams under low numerical aperture illumination \textit{Opt. Lett.} \textbf{36} 1605--7
\bibitem{Fickler(2014)} Fickler R, Lapkiewicz R, Ramelow S and Zeilinger A 2014 Quantum entanglement of complex photon polarization patterns in vector beams \textit{Phys. Rev. A} \textbf{89} 060301
\bibitem{Sivankutty(2016)} Sivankutty S, Andresen E R, Bouwmans G, Brown T G, Alonso M A and Rigneault H 2016 Single-shot polarimetry imaging of multicore fiber \textit{Opt. Lett.} \textbf{41} 2105--8
\bibitem{Reddy(2017)} Salla G R, Kumar V, Miyamoto Y and Singh R P 2017 Scattering of Poincar\'{e} beams: polarization speckles \textit{Opt. Express} \textbf{25} 19886--93
\bibitem{Garcia(2016)} Garcia--Gracia H and Guti\'{e}rrez--Vega J C 2016 Polarization singularities in nondiffracting Mathieu--Poincar\'{e} beams \textit{J. Opt.} \textbf{18} 014006
\bibitem{Herrero2006} Mart\'{i}nez-Herrero R, Mej\'{i}as P M and Piquero G 2006 Overall parameters for the characterization of non-uniformly totally polarized beams \textit{Opt. Commun.} \textbf{265} 6--10
\bibitem{Pitois2008} Pitois S, Fatome J and Millot G 2008 Polarization attraction using counter-propagating waves in optical fiber at telecommunication wavelengths \textit{Opt. Express} \textbf{16} 6646--51
\bibitem{Yan2005} Yan L, Yu Q and Willner A. E. 2005 Uniformly distributed states of polarization on the Poincar\'{e} sphere using an improved polarization scrambling scheme \textit{Opt. Commun.} \textbf{249} 43--50
\bibitem{Chipman2019} Chipman R A, Lam W T, Young G 2019 \textit{Polarized light and optical systems} (Boca Raton, FL: CRC Press)
\bibitem{Herrero(2008)} Mart\'{i}nez--Herrero R, Mej\'{i}as P M, Piquero G and Ram\'{i}rez--S\'{a}nchez V 2008 Global parameters for characterizing the radial and azimuthal polarization content of totally polarized beams \textit{Opt. Commun.} \textit{281} 1976--80
\bibitem{Plick2015} Plick W N and Krenn M 2015 Physical meaning of the radial index of Laguerre-Gauss beams \textit{Phys. Rev. A} \textbf{92} 063841
\bibitem{Zhan(2009)} Zhan Q 2009 Cylindrical vector beams: from mathematical concepts to applications \textit{Advances in Optics and Photonics} \textbf{1} 1--57
\bibitem{Otte2016} Otte E, Alpmann C, and Denz C 2016 Higher-order polarization singularitites in tailored vector beams \textit{J. Opt.} \textbf{18} 074012
\bibitem{Gutierrez2005} Guti\'{e}rrez-Vega J C and Bandres M A 2005 Helmholtz-Gauss waves \textit{J. Opt. Soc. Am. A} \textbf{22} 289--98
\bibitem{Durnin1987} Durnin J, Miceli J J and Eberly J H 1987 Diffraction-free beams \textit{Phys. Rev. Lett.} \textbf{58} 1499--501
\bibitem{Vaity2015} Vaity P and Rusch L 2015 Perfect vortex beam: Fourier transformation of a Bessel beam \textit{Opt. Lett.} \textbf{40} 597--600
\bibitem{Guasoni2014} Guasoni M, Ass\'{e}mat E, Morin P, Picozzi A, Fatome J, Pitois S, Jauslin H R, Millot T and Sugny D 2014 Line of polarization attraction in highly birefringent optical fibers \textit{J. Opt. Soc. Am. B} \textbf{31} 572--80
\bibitem{Moreno2016} Moreno I, Sanchez-Lopez M M, Badham K, Davis J A and Cottrell D M 2016 Generation of integer and fractional vector beams with q-plates encoded onto a spatial light modulator \textit{Opt. Lett.} \textbf{41} 1305--8
\bibitem{Liu2012} Liu S, Li P, Peng T and Zhao J 2012 Generation of arbitrary spatially variant polarization beams with a trapezoid Sagnac interferometer \textit{Opt. Express} \textbf{20} 21715--21.
\bibitem{Perez-Garcia2017} Perez-Garcia B, L\'{o}pez-Mariscal C, Hernandez-Aranda R I and Guti\'{e}rrez-Vega J C 2017 On-demand tailored vector beams \textit{App. Opt.} \textbf{56} 6967--72
\bibitem{Liu2018} Liu S, Qi S, Zhang Y, Li P, Wu D, Han L and Zhao J 2018 Highly efficient generation of arbitrary vector beams with tunable polarization, phase, and amplitude. \textit{Photonics Research} \textbf{6} 228--33
\bibitem{Yao2012} Yao L, Huang H, Chen J, Tan E and Willner A 2012 A novel scheme for achieving quasi-uniform rate polarization scrambling at 752 krad/s \textit{Opt. Express} \textbf{20} 1691--9
\end{thebibliography}
\end{document}